\documentclass[12pt]{article}
\usepackage{graphicx}

\newcommand{\beqa}{\begin{eqnarray}}
\newcommand{\eeqa}{\end{eqnarray}}
\newcommand{\beq}{\begin{equation}}
\newcommand{\eeq}{\end{equation}}
\newcommand{\non}{\nonumber}
\newcommand{\lab}{\label}

\newcommand{\svline}[1]{\multicolumn{1}{|c}{#1}}
\newcommand{\bra}{\langle}
\newcommand{\ket}{\rangle}
\newcommand{\qed}{\hbox{\rule[-2pt]{6pt}{6pt}}}

\begin{document}

\title
{Building partially entangled states
with Grover's amplitude amplification process}

\author{Hiroo Azuma\thanks{hiroo@crc.canon.co.jp}\\
Mathematical Engineering Division,\\
Canon Research Center,\\
5-1, Morinosato-Wakamiya, Atsugi-shi,\\
Kanagawa, 243-0193, Japan}

\date{January 12, 2000}

\maketitle

\begin{abstract}
We discuss how to build some partially entangled states
of $n$ two-state quantum systems (qubits).
The optimal partially entangled state
with a high degree of symmetry is
considered to be
useful for overcoming a shot noise limit
of Ramsey spectroscopy
under some decoherence.
This state is
invariant under permutation of any two qubits
and inversion
between the ground state $|0\ket$
and an excited state $|1\ket$
for each qubit.
We show that
using selective phase shifts in certain basis vectors
and Grover's inversion about average operations,
we can construct this high symmetric entangled state
by $(\mbox{polynomial in $n$})\times 2^{n/2}$
successive unitary transformations
that are applied on two or three qubits.
%%%%%%%%%%%%%%%%%%%%%%%%%%%%%%%%%%%%%%%%%
We can apply our method to build more general entangled states.
%%%%%%%%%%%%%%%%%%%%%%%%%%%%%%%%%%%%%%%%%
\end{abstract}

%%%%%%%%%%%%%%%%%%%%%%%%%%%%%%%%%%%%%%%%%
\section{Introduction}
\lab{INTRO}
Recently rapid progress
in quantum computation
and quantum information theory
have been made\cite{Feynman}\cite{Deutsch-Jozsa}.
In these fields, properties of quantum mechanics,
which are superposition,
interference,
and entanglement, are handled skillfully.
After Shor's algorithm for factorization
and discrete logarithms
and
Grover's algorithm for  search problems
appeared
\cite{Simon}\cite{Shor}\cite{Ekert-Jozsa}\cite{Grover}\cite{Boyer-Brassard},
many researchers have been proposing methods
for the realization of quantum computation and
developing quantum algorithms.
On the other hand, in the fields of quantum information theory,
it is recognized that entangled states play important roles
for robustness against decoherence\cite{Bennett-Fuchs}.

As an application of these results,
it is considered to overcome
the quantum shot noise limit
by using entangled states of
$n$ two-level systems (qubits)
for Ramsey
%%%%%%%%%%%%%%%%%%%%%%%%%%%%%%%%%%%%%%
\\
%%%%%%%%%%%%%%%%%%%%%%%%%%%%%%%%%%%%%%
spectroscopy\cite{Wineland92}\cite{Wineland94}.
(M.~Kitagawa et al.
gave a similar idea,
though an experimental scheme that they discussed
was not Ramsey spectroscopy
of qubits\cite{Kitagawa}.)
When we can neglect decoherence of the system
caused by an environment,
the maximally entangled state serves us
an improvement of a frequency measurement.
In this case,
the fluctuation of frequency is decreased by $1/\sqrt{n}$.
(In this paper, for example,
we consider
$(1/\sqrt{2})
(|0\cdots 0\ket +|1\cdots 1\ket)$
one of the maximally entangled states.
Entanglement for $n(\geq 3)$-qubit system
has not been defined clearly\cite{Bennett-DiVincenzo}.)
However,
if the decoherence is considered,
the maximally entangled state provides
the same resolution
that an uncorrelated system provides\cite{Huelga}.
S.~F.~Huelga et al.
proposed
using a partially entangled state
which has a high degree of symmetry.
This state is invariant under permutation of any two qubits
and inversion between the ground state $|0\ket$
and an excited state $|1\ket$ for each qubit.
If we prepare the high symmetric partially entangled state
optimized numerically,
it provides high resolution in comparison
with the maximally entangled states and uncorrelated states.

Carrying out an experiment of Ramsey spectroscopy
with the optimal high symmetric partially entangled state,
we have to prepare it for an initial state as soon as possible,
before time limit of decoherence.
In this paper,
we study how to construct this state efficiently.
We estimate time to prepare it
by the number of elementary quantum gates
that are unitary transformations
applied on two or three qubits
\cite{Shor}\cite{Ekert-Jozsa}\cite{Barenco}.
The number of gates is considered to be
in proportion to the amount of time for building the state.
We show it takes
$O((n^{3}\log_{2}n)\times 2^{n/2})$ steps at most
to build it.
(It was shown that any unitary transformation
$U$ $(\in \mbox{\boldmath $U$}(2^{n}))$
can be constructed from $O(n^{3}2^{2n})$ elementary gates
at most\cite{Barenco}.)
Furthermore, our method can be applied to build
more general entangled states.

Before discussing how to build partially entangled states,
we try to construct the maximally entangled state
with $n$ qubits from an initial state $|0\cdots 0\ket$.
To do it,
we need two unitary transformations for elementary gates.
They are $H^{(j)}$ (the Walsh-Hadamard transformation)
which operates on the $j$-th qubit,
and
$\bigwedge_{1}^{(j,k)}(\sigma_{x})$
which operates on the $j$-th and $k$-th qubits:
\[
H^{(j)}=\frac{1}{\sqrt{2}}
\begin{array}{ccc}
\bra 0|      & \bra 1| &\\
\svline{1} & 1         & \svline{|0\ket}\\
\svline{1} & -1       & \svline{|1\ket}
\end{array},
\quad\quad
\mbox{$\bigwedge$}_{1}^{(j,k)}(\sigma_{x})=
\begin{array}{ccccc}
\bra 00|     & \bra 01| & \bra 10| & \bra 11| &\\
\svline{1}  & 0           & 0            & 0           & \svline{|00\ket}\\
\svline{0}  & 1           & 0            & 0           & \svline{|01\ket}\\
\svline{0}  & 0           & 0            & 1           & \svline{|10\ket}\\
\svline{0}  & 0           & 1            & 0           & \svline{|11\ket}
\end{array}.
\]
Because $\bigwedge_{1}^{(j,k)}(\sigma_{x})$
transforms $|x,y\ket$ $(x,y\in \{0,1\})$
to $|x,x\oplus y\ket$
(applying $\sigma_{x}$ on $k$-th qubit
according to the $j$-th qubit),
it is sometimes called the controlled-NOT gate.
Applying
$\bigwedge_{1}^{(1,n)}(\sigma_{x})\cdots
\bigwedge_{1}^{(1,2)}(\sigma_{x})H^{(1)}$
on
$|0\ket_{1}\otimes\cdots\otimes |0\ket_{n}$,
we can obtain the maximally entangled state,
$(1/\sqrt{2})(|0\cdots 0\ket + |1\cdots 1\ket)$.

But, building partially entangled states
like
\beqa
|\psi_{4}\ket
&=&a_{0}|0\ket_{s}+a_{1}|1\ket_{s}+a_{2}|2\ket_{s} \non \\
&\equiv&a_{0}(|0000\ket +|1111\ket ) \non \\
&&\quad
+a_{1}(|0001\ket +|0010\ket +|0100\ket+|1000\ket \non \\
&&\quad\quad\quad\quad  
+|1110\ket +|1101\ket +|1011\ket+|0111\ket ) \non \\
&&\quad
+a_{2}(|0011\ket +|0101\ket +|0110\ket+|1001\ket  
+|1010\ket +|1100\ket )
\lab{Psi4Form}
\eeqa
(this is an example of the $4$-qubit high symmetric
partially entangled state),
where $a_{0}$, $a_{1}$, $a_{2}$ are given (real)
coefficients,
and $|k\ket_{s}$ is an equally weighted superposition of
$k$ or $(4-k)$ excited qubits,
we feel difficult.
It is hard to resolve a unitary transformation
that transforms $|0000\ket$ to $|\psi_{4}\ket$
into local operations
like $H^{(j)}$ or $\bigwedge_{1}^{(j,k)}(\sigma_{x})$.
This is because we don't know a systematic method
for adjusting coefficients of basis vectors.
This matter is a motivation of this paper.

This paper is arranged as follows.
In \S\ref{HiSymParEntSta},
we explicitly describe
the high symmetric partially entangled states.
We make preparations for our method of building them.
In \S\ref{MakMasVecBeWeiEq},
we introduce a unitary transformation
that makes two sets of basis vectors
classified by their coefficients
be weighted equally.
We derive a sufficient condition
for finding an appropriate parameter
that characterizes this transformation.
In \S\ref{CaseWhSufCondNotSat},
we develop a technique which transforms
the state that doesn't satisfy the sufficient
condition derived in \S\ref{MakMasVecBeWeiEq}
into a state
that satisfies it.
This technique is an application of
Grover's amplitude amplification process\cite{Grover}\cite{Boyer-Brassard}.
In \S\ref{WholeProc},
we show the whole procedure for building
the high symmetric entangled states
and give a sketch of
implementation for it.
We estimate the whole number of elementary gates
of our method.
%%%%%%%%%%%%%%%%%%%%%%%%%%%%%%%%%%%%%%%%%%%%%%
We also show that we can use our procedure
for building more general entangled states.
%%%%%%%%%%%%%%%%%%%%%%%%%%%%%%%%%%%%%%%%%%%%%%
%
In \S\ref{DISCUSSION},
we give a brief discussion.
%%%%%%%%%%%%%%%%%%%%%%%%%%%%%%%%%%%%%%%%%%%%%%
In {\S}Appendix,
we construct networks of quantum gates
for our method concretely,
and derive a variation of coefficients
of the state under the transformation
discussed in \S\ref{CaseWhSufCondNotSat}.
%%%%%%%%%%%%%%%%%%%%%%%%%%%%%%%%%%%%%%%%%%%%%%

%%%%%%%%%%%%%%%%%%%%%%%%%%%%%%%%%%%%%%%%%
\section{High symmetric partially entangled states}
\lab{HiSymParEntSta}
In this section,
we define high symmetric partially entangled states
explicitly.
We also make preparations
for our method of building them,
defining an initial state,
giving some unitary transformations
used frequently,
and so on.

The partially entangled state
which has a high degree of symmetry
is given by
\begin{equation}
|\psi_{n}\ket=\sum_{k=0}^{\lfloor n/2 \rfloor} a_{k}|k\ket_{s}
\quad\quad
\mbox{for $n\geq 2$},
\lab{PsiNForm}
\end{equation}
where
$\lfloor n/2 \rfloor$ is the maximum integer
that doesn't exceed $n/2$\cite{Huelga}.
$\{a_{k}\}$ are given real coefficients.
We assume $a_{k}\geq 0$ for
$k=0,\cdots,\lfloor n/2 \rfloor$
for a while.
$|k\ket_{s}$ is an equally weighted superposition of
$k$ or $(n-k)$ excited qubits,
as shown in
$|\psi_{4}\ket$ of Eq.~(\ref{Psi4Form}).
This state has symmetric properties,
invariance under permutation of any two qubits,
and
invariance under inversion between $|0\ket$ and $|1\ket$
for each qubit.
A main aim of this paper is
to show a procedure for building
$|\psi_{n}\ket$ efficiently.
We emphasize that
$\{a_{k}\}$ of $|\psi_{n}\ket$ in
Eq.~(\ref{PsiNForm}) are given
and numerically optimized to
realize high precision for Ramsey spectroscopy.

We make some preparations.
To build $|\psi_{n}\ket$,
we prepare an $n$-qubit register in a uniform superposition
of $2^{n}$ binary states,
$(1/\sqrt{2^{n}})\sum_{x\in\{0,1\}^{n}}|x\ket$
($\{0,1\}^{n}$ represents
a set of all $n$-bit binary strings),
and apply unitary transformations on the register successively.
(Initializing the register to $|0\cdots 0\ket$ and
applying $H^{(j)}$ $(1\leq j \leq n)$ on each qubit,
we can obtain the uniform superposition.)

In our method,
we use two kinds of transformations.
One of them is a selective phase shift transformation
in certain basis vectors.
It is given by the
$2^{n}\times 2^{n}$ diagonal matrix form,
\begin{equation}
R_{xy}=
\left\{\begin{array}{ll}
\exp(i\theta_{x}), & \mbox{for $x=y$} \\
0,                 & \mbox{for $x\neq y$}
\end{array}
\right.,
\lab{matrix-phaseshift}
\end{equation}
where subscripts $x$, $y$ represent the basis
vectors $\{|x\ket|x\in\{0,1\}^{n}\}$
and $0\leq\theta_{x} <2\pi$ for $\forall x$.
(Although a general phase shift
transformation in the form of Eq.~(\ref{matrix-phaseshift})
takes a number of elementary gates
exponential in $n$ at most,
we use only special transformations
that need polynomial steps.
This matter is discussed
in \S\ref{WholeProc}
%%%%%%%%%%%%%%%%%%%%%%%%%%%%%%%
and {\S}Appendix~A.)
%%%%%%%%%%%%%%%%%%%%%%%%%%%%%%%
The other is
Grover's inversion about average operation
$D$\cite{Grover}.
The $2^{n}\times 2^{n}$ matrix representation of
$D$ is given by
\begin{equation}
D_{xy}=
\left\{\begin{array}{ll}
-1+ 2^{-n+1},    & \mbox{for $x=y$} \\
2^{-n+1},        & \mbox{for $x\neq y$}
\end{array}
\right..
\lab{matrix-GroversD}
\end{equation}

Because we use only unitary transformations
and never measure any qubits,
we can regard our procedure
for building $|\psi_{n}\ket$
as a succession of unitary transformations.
For simplicity,
we consider a chain of transformations reversely
to be a
transformation from $|\psi_{n}\ket$ to the uniform superposition
instead of it from the uniform superposition
to $|\psi_{n}\ket$.
Fortunately, an inverse operation of
the selective phase shift on certain basis vectors
is also the phase shift,
and an inverse operation of $D$ defined in Eq.~(\ref{matrix-GroversD})
is also $D$.
In the rest of this paper,
because of simplicity,
we describe the procedure
reversely from $|\psi_{n}\ket$ to the uniform superposition.
(Building $|\psi_{n}\ket$ actually,
we carry out the inversion of the procedure.)

%%%%%%%%%%%%%%%%%%%%%%%%%%%%%%%%%%%%%%%%%
\section{Making basis vectors
be weighted equally}
\lab{MakMasVecBeWeiEq}
At first, we show how to transform $|\psi_{2}\ket$
to the uniform superposition as an example.
After that,
we consider a case of $|\psi_{n}\ket$ for $n\geq 3$.

Writing $|\psi_{2}\ket$ as
\[
|\psi_{2}\ket
=
a_{0}(|00\ket+|11\ket)
+a_{1}(|01\ket+|10\ket)
\quad\quad
\mbox{where $a_{0}\geq0$, $a_{1}\geq0$
and
$a_{0}^{2}+a_{1}^{2}=1/2$},
\]
we apply the following transformations on it.
Shifting the phase of $|01\ket$ by $\theta$
and
shifting the phase of $|10\ket$ by $(-\theta)$,
we obtain
\[
a_{0}(|00\ket+|11\ket)
+a_{1}e^{i\theta}|01\ket
+a_{1}e^{-i\theta}|10\ket.
\]
The value of $\theta$ is considered later.
Then, we apply $D$ on the above state.
$D$ is given as
\[
D=\frac{1}{2}
\begin{array}{ccccc}
\bra 00|     & \bra 01| & \bra 10| & \bra 11| &\\
\svline{-1}  & 1           & 1            & 1           & \svline{|00\ket}\\
\svline{1}   & -1          & 1            & 1           & \svline{|01\ket}\\
\svline{1}   & 1           & -1           & 1           & \svline{|10\ket}\\
\svline{1}   & 1           & 1            & -1          & \svline{|11\ket}
\end{array},
\]
and we get
\[
A_{0}(|00\ket+|11\ket)
+A_{1}|01\ket
+A_{1}^{*}|10\ket
\quad\quad
\mbox{where $A_{0}=a_{1}\cos\theta$,
$A_{1}=a_{0}-ia_{1}\sin\theta$}.
\]
Defining $\phi$ as $e^{i\phi}\equiv A_{1}/|A_{1}|$,
we shift the phase of $|01\ket$ by $(-\phi)$ and
shift the phase of $|10\ket$ by $\phi$.
We get
\[
A_{0}(|00\ket+|11\ket)
+|A_{1}|(|01\ket+|10\ket).
\]
If $A_{0}=|A_{1}|$,
we obtain the uniform superposition.
Here,
we can assume
$0\leq a_{0}<1/2<a_{1}$
without losing generality.
From these considerations,
the value of $\theta$ is given by $\cos\theta=1/(2a_{1})$.

In case of $n\geq 3$,
we take the following method.
Classifying basis vectors
$\{|x\ket|x\in\{0,1\}^{n}\}$
of $|\psi_{n}\ket$
by their coefficients,
we obtain $(\lfloor n/2 \rfloor +1)$ sets of them
characterized by $a_{k}$.
We consider the transformation that makes
two sets of basis vectors
(e.g. sets of basis vectors with $a_{0}$ and $a_{1}$)
be weighted equally and reduces the number of sets by one.
If we do this operation for $\lfloor n/2 \rfloor$ times,
we obtain the uniform superposition.

Here, we consider how to make
a set of basis vectors with $a_{1}$ be weighted equally
to a set of them with $a_{0}$ on $|\psi_{n}\ket$.
A similar discussion can be applied on other sets of them.
From now,
we write $|\psi_{n}\ket$ as
\begin{equation}
|\Psi\ket
=[
\underbrace{a_{0},\cdots,}_{2l}
\underbrace{a_{1},\cdots,}_{2m}
a_{2(l+m)},\cdots,a_{2^{n}-1}
]
\quad\quad
\mbox{for $n\geq 2$}.
\lab{procedure-step2}
\end{equation}
As the representation of Eq.~(\ref{procedure-step2}),
we sometimes write a column vector by a row vector.
In Eq.~(\ref{procedure-step2}),
we order the orthonormal basis vectors
$\{|x\ket|x\in\{0,1\}^{n}\}$
appropriately,
and
coefficients $a_{0}$ and $a_{1}$ are put
in the left side of the row.
Because $|\psi_{n}\ket$ is
invariant under inversion between $|0\ket$
and $|1\ket$ for each qubit,
the number of basis vectors that have a coefficient
$a_{k}$ $(0\leq k\leq \lfloor n/2 \rfloor)$
is even.
Therefore, we can give the number of $a_{0}$
by $2l$ and the number of $a_{1}$
by $2m$,
where $l\geq 1$, $m\geq 1$, and $l+m\leq 2^{n-1}$.
The other $(2^{n}-2l-2m)$ coefficients,
$\{a_{2},\cdots,a_{\lfloor n/2 \rfloor}\}$,
are gathered in the right side of the row
and they are relabeled
$\{a_{j}|2(l+m)\leq j \leq 2^{n}-1\}$.
Reordering basis vectors
never changes matrix forms of
$R$ defined in Eq.~(\ref{matrix-phaseshift})
and $D$ defined in Eq.~(\ref{matrix-GroversD}),
except for permutation of diagonal elements
of $R$.

We carry out the following transformations.
Firstly,
we shift phases of $m$ basis vectors
with coefficients $a_{1}$
by $\theta$
and
shift phases of the other $m$ basis vectors
with coefficients $a_{1}$
by $(-\theta)$.
How to choose the value of $\theta$ is discussed later.
We obtain
\begin{equation}
R_{\theta}|\Psi\ket=
[
\underbrace{a_{0},\cdots,}_{\mbox{$2l$}}
\underbrace{e^{i\theta}a_{1},\cdots,}_{\mbox{$m$}}
\underbrace{e^{-i\theta}a_{1},\cdots,}_{\mbox{$m$}}
a_{2(l+m)},\cdots,a_{2^{n}-1}
],
\end{equation}
where $0\leq\theta<2\pi$
($R_{\theta}$ is given by $2^{n}\times 2^{n}$ diagonal matrix
whose diagonal elements are
$\{1,\cdots,e^{i\theta},\cdots,e^{-i\theta},\cdots,1,\cdots,1\}$
).

Then we apply $D$ on $R_{\theta}|\Psi\ket$,
\begin{equation}
DR_{\theta}|\Psi\ket=
[
\underbrace{A_{0},\cdots,}_{\mbox{$2l$}}
\underbrace{A_{1},\cdots,}_{\mbox{$m$}}
\underbrace{A_{1}^{*},\cdots,}_{\mbox{$m$}}
A_{2(l+m)},\cdots,A_{2^{n}-1}
],
\end{equation}
where
\begin{equation}
\left\{
\begin{array}{rcl}
2^{n-1}A_{0}&=&(2l-2^{n-1})a_{0}+2ma_{1}\cos\theta+C, \\
2^{n-1}A_{1}
&=&2la_{0}+(m-2^{n-1})a_{1}e^{i\theta}
+ma_{1}e^{-i\theta}+C, \\
2^{n-1}A_{j}
&=&2la_{0}+2ma_{1}\cos\theta-2^{n-1}a_{j}+C,
\end{array}
\right.
\lab{A-Coefficients}
\end{equation}
for $j=2(l+m),\cdots,2^{n}-1$,
and
$C=\sum_{j=2(l+m)}^{2^{n}-1}a_{j}$.
We notice that $A_{i}=A_{j}$,
if $a_{i}=a_{j}$
for $2(l+m)\leq \forall i,j \leq 2^{n}-1$.

Finally, we apply the selective phase shift
to cancel the phases of $A_{1}$ and $A_{1}^{*}$.
Defining $\phi$ as
$e^{i\phi}=A_{1}/|A_{1}|$,
we shift the phases of $m$ basis vectors
with coefficients $A_{1}$
by $(-\phi)$
and
shift the phases of $m$ basis vectors
with coefficients $A_{1}^{*}$
by $\phi$.
We obtain
\begin{equation}
\tilde{R}_{\theta}DR_{\theta}|\Psi\ket=
[
\underbrace{A_{0},\cdots,}_{\mbox{$2l$}}
\underbrace{|A_{1}|,\cdots,}_{\mbox{$2m$}}
A_{2(l+m)},\cdots,A_{2^{n}-1}
].
\lab{RDR-column}
\end{equation}
We write the second phase shift operator
as $\tilde{R}_{\theta}$,
because
the phase shift angle $\phi$
depends on $\theta$ and $\{a_{k}\}$.

If we can choose $\theta$ to let $|A_{1}|$ be equal to $A_{0}$,
we succeed in making two sets of basis vectors
characterized by $a_{0}$ and $a_{1}$ be weighted equally.
From now,
we call this series of operations an $(\tilde{R}DR)$ operation.
If we can carry out the $(\tilde{R}DR)$ operations,
with suitable parameters $\theta$s,
$\lfloor n/2 \rfloor$ times on $|\psi_{n}\ket$,
we get the uniform superposition.

However, there are two difficulties.
We can't always find
a suitable $\theta$ that lets $|A_{1}|$ be equal to $A_{0}$
for the $(\tilde{R}DR)$ operation
on an arbitrary given $|\psi_{n}\ket$.
We consider the next lemma
that
shows a sufficient condition
for finding a suitable $\theta$.
It gives us a hint
which couple of sets of basis vectors do we let be weighted equally.

\vspace*{12pt}
\noindent
{\bf Lemma~1:}
We define an $n$-qubit state
$|\Psi\ket$ as
\begin{equation}
|\Psi\ket=
[\underbrace{a_{0},\cdots,}_{\mbox{$2l$}}
\underbrace{a_{1},\cdots,}_{\mbox{$2m$}}
a_{2(l+m)},\cdots,a_{2^{n}-1}]
\quad\quad
\mbox{for $n\geq 2$},
\lab{ReductionLemma1-Psi}
\end{equation}
where $0\leq a_{j}$ for $j=0,1,2(l+m),\cdots,2^{n}-1$
and
$a_{0}<a_{1}$.
The basis vectors of Eq.~(\ref{ReductionLemma1-Psi})
are $\{|x\ket|x\in\{0,1\}^{n}\}$.
We assume that 
the number of elements $a_{0}$ is equal to $2l$
and
the number of elements $a_{1}$ is equal to $2m$,
where $l\geq 1$, $m\geq 1$ and $l+m\leq 2^{n-1}$.
We write a sum of all coefficients by
\begin{equation}
S
=2la_{0}+2ma_{1}+\sum_{j=2(l+m)}^{2^{n}-1}a_{j}.
\end{equation}
If the following condition is satisfied,
\begin{equation}
S-2^{n-2}(a_{0}+a_{1})\geq 0,
\lab{suff-cond-Lemma1}
\end{equation}
we can always make $2(l+m)$ basis vectors
whose coefficients are $a_{0}$ or $a_{1}$
be weighted equally by the $(\tilde{R}DR)$ operation
in which $R$ and $\tilde{R}$ are applied on
$2m$ basis vectors with $a_{1}$.

\vspace*{12pt}
\noindent
{\bf Proof:}
$\tilde{R}_{\theta}DR_{\theta}|\Psi\ket$
is given by Eq.~(\ref{A-Coefficients}) and Eq.~(\ref{RDR-column}).
To evaluate a difference between
$A_{0}^{2}$ and $|A_{1}|^{2}$,
we define
\beqa
f(\theta)
&=&2^{n-2}(A_{0}^{2}-|A_{1}|^{2}) \non \\
&=&
(2la_{0}+2ma_{1}\cos\theta+C)(a_{1}\cos\theta-a_{0})
-2^{n-2}(a_{1}^{2}-a_{0}^{2}).
\lab{FunctionFTheta}
\eeqa
If $f(\theta)=0$,
$A_{0}^{2}$ is equal to $|A_{1}|^{2}$.
We estimate $f(0)$ and $f(\pi/2)$,
\begin{equation}
\left\{
\begin{array}{rcl}
f(0)&=&(a_{1}-a_{0})[S-2^{n-2}(a_{0}+a_{1})], \\
f(\pi/2)&=&-a_{0}(2la_{0}+C)-2^{n-2}(a_{1}^{2}-a_{0}^{2})<0.
\end{array}
\right.
\end{equation}
If $S-2^{n-2}(a_{0}+a_{1})\geq 0$,
there is $0\leq \theta <(\pi/2)$,
which satisfies $A_{0}^{2}=|A_{1}|^{2}$.
If signs of $A_{0}$ and $|A_{1}|$ are different from each other,
the phase shift by $\pi$ on basis vectors
with negative coefficients is done.
\hfill\qed

To find the suitable sequence of sets of basis vectors
that we make be weighted equally,
we take the following procedure.
(For $n=2,3$, the condition of Eq.~(\ref{suff-cond-Lemma1})
is always satisfied.
Therefore, we consider the case of $n\geq 4$.)
We describe a given state $|\psi_{n}\ket$ by
Eq.~(\ref{PsiNForm}),
where
$n\geq 4$ and
$a_{k}\geq 0$ for $0\leq k\leq \lfloor n/2 \rfloor$.
Let $a_{min}$  be the minimum coefficient
among $\{a_{k}\}$
and
$a_{min+1}$ be the coefficient next to $a_{min}$
($0\leq a_{min}<a_{min+1}<a_{j}$,
where $a_{j}$ is any coefficient of $|\psi_{n}\ket$
except $a_{min}$ and $a_{min+1}$).
Because the number of different coefficients
in $\{a_{k}\}$ is
equal to $(\lfloor n/2 \rfloor+1)$,
it takes $O(n)$ steps to find $a_{min}$ and $a_{min+1}$
on classical computation.

\begin{enumerate}
\item
If $S<2^{n-2}(a_{min}+a_{min+1})$,
we get $S<2^{n-2}(a_{i}+a_{j})$ for $\forall i,j$.
In this case,
it can't be guaranteed
to find a good $\theta$ for the $(\tilde{R}DR)$ operation.
We take another technique explained in the next section.
\item
If $S\geq 2^{n-2}(a_{min}+a_{min+1})$,
we can find a good $\theta$ for
the $(\tilde{R}DR)$ operation
and get a relation,
$A_{min}^{2}=|A_{min+1}|^{2}$.
Because Eq.~(\ref{FunctionFTheta})
is an equation of the second degree
for $\cos\theta$,
we can obtain $\theta$ with some calculations.
In this case,
though we may make
other couples of sets of basis vectors be weighted equally,
we neglect them.
Shifting phases of basis vectors which have negative coefficients
by $\pi$ after the $(\tilde{R}DR)$ operation,
we obtain a state whose all coefficients are
nonnegative.
There are
$\lfloor n/2 \rfloor$ kinds of new coefficients in the state after
these operations,
and we can derive them from
Eq.~(\ref{A-Coefficients}) and Eq.~(\ref{RDR-column})
with $poly(n)$ steps by classical computation
($poly(n)$ means polynomial in $n$).
We can check whether the condition of
Lemma~1
is satisfied or not again.
\end {enumerate}

%%%%%%%%%%%%%%%%%%%%%%%%%%%%%%%%%%%%%%%%%
\section{The case where the sufficient condition
isn't satisfied}
\lab{CaseWhSufCondNotSat}
In this section,
we consider how to make a couple of sets of
basis vectors be weighted equally
in the case where the state doesn't
satisfy the condition of Lemma~1,
$S\geq 2^{n-2}(a_{min}+a_{min+1})$.
We develop a technique that adjusts
amplitudes of basis vectors
and transforms the state
to a state that satisfies the sufficient condition.
This is an application of Grover's
amplitude amplification process.

For example, we consider the state
which has two kinds of coefficients,
\begin{equation}
|\Psi\ket
=[\underbrace{a_{0},\cdots,}_{\mbox{$(2^{n}-t)$}}
\underbrace{a_{1},\cdots}_{\mbox{$t$}}]
\quad\quad
\mbox{for $n\geq 4$},
\mbox{where $0\leq a_{0}<a_{1}$}.
\lab{SimpleModelForRpiDIteration}
\end{equation}
We assume that 
the number of elements $a_{1}$ is equal to $t$,
where $2\leq t\leq2^{n}-2$, and $t$ is even.
If $0<t<2^{n-2}$ and
$[(3\cdot 2^{n-2}-t)/(2^{n-2}-t)]a_{0}<a_{1}$
($a_{1}$ is bigger enough than $a_{0}$),
we obtain $S<2^{n-2}(a_{0}+a_{1})$
for $|\Psi\ket$.

\begin{figure}[ht]
\begin{center}
\includegraphics[scale=0.8]{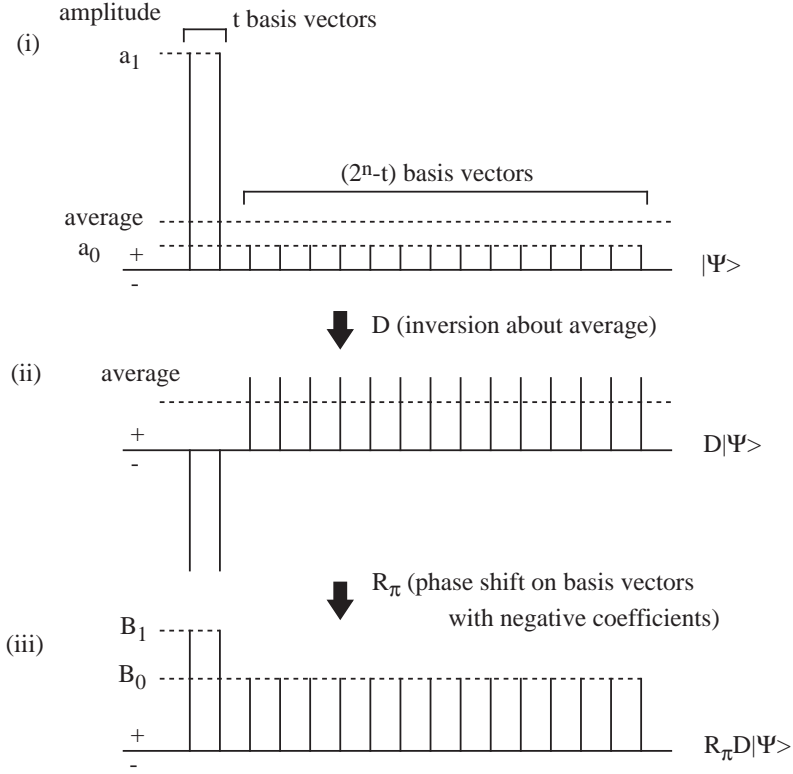}
\caption{A variation of coefficients
under the $(R_{\pi}D)$ transformation.}
\label{varco2-epsf}
\end{center}
\end{figure}
In this case,
applying $D$
(we use the property of
the inversion about average operation),
and then,
shifting the phase by $\pi$ on basis vectors
which have negative coefficients,
we can reduce a difference between
new coefficients, $B_{0}$ and $B_{1}$,
as shown in
Figure~\ref{varco2-epsf}.
We write this phase shift operation by $R_{\pi}$.
It can be expected that
$[S-2^{n-2}(a_{min}+a_{min+1})]$ gets bigger
by applying $(R_{\pi}D)$ successively.
The next lemma shows it clearly.

\vspace*{12pt}
\noindent
{\bf Lemma~2:}
We consider a state,
\begin{equation}
|\Psi\ket=
[\underbrace{a_{0},\cdots,}_{\mbox{$2l$}}
\underbrace{a_{1},\cdots,}_{\mbox{$2m$}}
a_{2(l+m)},\cdots,a_{2^{n}-1}]
\quad\quad
\mbox{for $n\geq 4$},
\end{equation}
where $0\leq a_{0}<a_{1}<a_{j}$
for $j=2(l+m),\cdots,2^{n}-1$.
We assume that 
the number of elements $a_{0}$ is equal to $2l$
and
the number of elements $a_{1}$ is equal to $2m$,
where $l\geq 1$, $m\geq 1$, and $l+m\leq 2^{n-1}$.
We also assume $S$,
a sum of all coefficients of $|\Psi\ket$,
has the relation,
\begin{equation}
S-2^{n-2}(a_{0}+a_{1})<0.
\lab{AssumptionOfLemma2}
\end{equation}
Applying the inversion about average operation $D$ on $|\Psi\ket$,
and then,
doing the phase shift transformation by $\pi$ on basis vectors
which have negative coefficients,
we obtain
\begin{equation}
R_{\pi}D|\Psi\ket=
[B_{0},\cdots,
B_{1},\cdots,
B_{2(l+m)},\cdots,B_{2^{n}-1}].
\lab{RpiDPsi-form-Lemma2}
\end{equation}
We define $\tilde{S}$ as a sum of all coefficients
of $R_{\pi}D|\Psi\ket$.
We also define
\begin{equation}
\left\{
\begin{array}{rcl}
\epsilon^{(0)}&=&(2l-2^{n-1})a_{0}+(2^{n}-2l)a_{1}, \non \\
\epsilon^{(1)}&=&(2l-2^{n-1})B_{0}+(2^{n}-2l)B_{1}.
\end{array}
\right.
\lab{def-epsilon0and1}
\end{equation}

\begin{enumerate}
\item
We get
$0<B_{0}<B_{1}<B_{j}$ for $j=2(l+m),\cdots,2^{n}-1$
and
\begin{equation}
[\tilde{S}-2^{n-2}(B_{0}+B_{1})]
-[S-2^{n-2}(a_{0}+a_{1})]
>\epsilon^{(0)}>0.
\end{equation}
\item
We obtain the relation,
\begin{equation}
\epsilon^{(1)}-\epsilon^{(0)}
\geq
\frac{2^{n}-2l}{2^{n-2}}
[2^{n-2}(a_{0}+a_{1})-S]>0.
\end{equation}
\end{enumerate}

\vspace*{12pt}
\noindent
{\bf Proof:}
We can derive
$D|\Psi\ket=
[a'_{0},\cdots,
a'_{1},\cdots,
a'_{2(l+m)},\cdots,a'_{2^{n}-1}]$,
where
\begin{equation}
\left\{
\begin{array}{rcl}
2^{n-1}a'_{0}&=&S-2^{n-1}a_{0}, \\
2^{n-1}a'_{1}&=&S-2^{n-1}a_{1}, \\
2^{n-1}a'_{j}&=&S-2^{n-1}a_{j},
\quad
\mbox{for $2(l+m)\leq j\leq 2^{n}-1$}.
\end{array}
\right.
\lab{a-coefficient}
\end{equation}
It is clear that $S-2^{n-1}a_{0}>0$.
Using the assumption of Eq.~(\ref{AssumptionOfLemma2}),
we obtain
$S-2^{n-1}a_{k}<0$ for $\forall k\neq 0$.
Therefore, we get $R_{\pi}D|\Psi\ket$
of Eq.~(\ref{RpiDPsi-form-Lemma2}),
where
\begin{equation}
\left\{
\begin{array}{rcl}
2^{n-1}B_{0}&=&S-2^{n-1}a_{0}, \\
2^{n-1}B_{1}&=&-S+2^{n-1}a_{1}, \\
2^{n-1}B_{j}&=&-S+2^{n-1}a_{j},
\quad
\mbox{for $2(l+m)\leq j\leq 2^{n}-1$}.
\end{array}
\right.
\lab{CoefficientsB1}
\end{equation}
We can derive a difference of $B_{1}$ and $B_{0}$
with the assumption of Eq.~(\ref{AssumptionOfLemma2}),
\begin{equation}
2^{n-1}(B_{1}-B_{0})=-2[S-2^{n-2}(a_{0}+a_{1})]>0.
\end{equation}
It is clear that
$B_{1}<B_{j}$ for $j=2(l+m),\cdots,2^{n}-1$.
We obtain the relation,
$0<B_{0}<B_{1}<B_{j}$ for $j=2(l+m),\cdots,2^{n}-1$.

Since
\[
\tilde{S}=\frac{4l}{2^{n-1}}(S-2^{n-1}a_{0})-S,
\quad\quad
\mbox{and}
\quad\quad
B_{0}+B_{1}=a_{1}-a_{0},
\]
we can derive $\Delta$
that is a variation 
of $[S-2^{n-2}(a_{0}+a_{1})]$
caused by the $(R_{\pi}D)$ operation,
\beqa
\Delta
&=&[\tilde{S}-2^{n-2}(B_{0}+B_{1})]
-[S-2^{n-2}(a_{0}+a_{1})] \non \\
&=&2(\frac{2l}{2^{n-1}}-1)S
-(4l-2^{n-1})a_{0}.
\lab{DeltaRepresentation1}
\eeqa

To estimate $\Delta$ precisely,
we prepare some useful relations.
From the definition of $S$,
we get
\begin{equation}
S=2la_{0}+2ma_{1}+\sum_{j=2(l+m)}^{2^{n}-1}a_{j}
\geq 2la_{0}+(2^{n}-2l)a_{1}.
\lab{ConditionOfS1}
\end{equation}
Using the assumption of Eq.~(\ref{AssumptionOfLemma2})
and Eq.~(\ref{ConditionOfS1}),
we can derive the relation,
\beqa
0
&>&S-2^{n-2}(a_{0}+a_{1}) \non \\
&\geq&2la_{0}+(2^{n}-2l)a_{1}-2^{n-2}(a_{0}+a_{1}) \non \\
&=&(2l-2^{n-2})a_{0}+(3\cdot 2^{n-2}-2l)a_{1}.
\lab{StrictUnequation1}
\eeqa
We modify the relation of Eq.~(\ref{StrictUnequation1})
and get a rougher relation,
\begin{equation}
0
>2la_{0}-2^{n-2}a_{1}+(3\cdot 2^{n-2}-2l)a_{1}
=2la_{0}+(2^{n-1}-2l)a_{1}.
\end{equation}
Because $0\leq a_{0}<a_{1}$,
we obtain
$2l-2^{n-1}>0$.
Seeing this relation and Eq.~(\ref{StrictUnequation1}) again,
we also obtain
\begin{equation}
2l>3\cdot 2^{n-2}.
\lab{2lCondition2}
\end{equation}

Here, we can estimate $\Delta$.
Because of Eq.~(\ref{2lCondition2}),
we can substitute Eq.~(\ref{ConditionOfS1})
for Eq.~(\ref{DeltaRepresentation1}),
\beqa
\Delta
&\geq&2(\frac{2l}{2^{n-1}}-1)[2la_{0}+(2^{n}-2l)a_{1}]
-(4l-2^{n-1})a_{0} \non \\
&=&
\frac{1}{2^{n-1}}(4l-3\cdot 2^{n-1})
[(2l-2^{n-2})a_{0}
+(3\cdot 2^{n-2}-2l)a_{1}] \non \\
&&\quad\quad\quad\quad\quad\quad
+2^{n-2}(a_{1}-a_{0}).
\label{Delta-ineqality}
\eeqa
Seeing Eq.~(\ref{2lCondition2}),
we find $3\cdot 2^{n-2}<2l<2^{n}$.
Therefore, we can derive the relation,
$0<(4l-3\cdot 2^{n-1})<2^{n-1}$.
From Eq.~(\ref{StrictUnequation1})
and Eq.~(\ref{Delta-ineqality}),
we can estimate $\Delta$,
\begin{equation}
\Delta
>
[(2l-2^{n-2})a_{0}
+(3\cdot 2^{n-2}-2l)a_{1}]
+2^{n-2}(a_{1}-a_{0})
=\epsilon^{(0)}
>0.
\end{equation}
The first result is derived.

From the definition
of Eq.~(\ref{def-epsilon0and1})
and Eq.~(\ref{CoefficientsB1}),
(\ref{ConditionOfS1}),
(\ref{StrictUnequation1}),
we can estimate the difference
between $\epsilon^{(0)}$ and $\epsilon^{(1)}$,
\beqa
\epsilon^{(1)}-\epsilon^{(0)}
&=&
\frac{1}{2^{n-1}}
[(4l-3\cdot 2^{n-1})S-2^{n}a_{0}(2l-2^{n-1})] \non \\
&\geq&
\frac{1}{2^{n-1}}
\{(4l-3\cdot 2^{n-1})[2la_{0}+(2^{n}-2l)a_{1}]
-(2l-2^{n-1})2^{n}a_{0}\} \non \\
&=&
-\frac{2^{n}-2l}{2^{n-2}}
[(3\cdot 2^{n-2}-2l)a_{1}
+(2l-2^{n-2})a_{0}] \non \\
&\geq&
\frac{2^{n}-2l}{2^{n-2}}
[2^{n-2}(a_{0}+a_{1})-S]>0.
\eeqa
The second result is derived.
\hfill\qed

Because of Lemma~2,
doing the $(R_{\pi}D)$ transformations successively,
we can make
$[S-2^{n-2}(a_{0}+a_{1})]$ be nonnegative.
We explain this matter as follows.
We consider the state $|\Psi^{(0)}\ket$
specified with coefficients,
$0\leq a_{0}<a_{1}<a_{j}$
for $2(l+m)\leq j\leq 2^{n}-1$,
and assume
$S-2^{n-2}(a_{0}+a_{1})<0$.
We apply $(R_{\pi}D)$ on $|\Psi^{(0)}\ket$
and obtain $|\Psi^{(1)}\ket$
described with coefficients,
$0<B_{0}^{(1)}<B_{1}^{(1)}<B_{j}^{(1)}$
for $2(l+m)\leq j\leq 2^{n}-1$.
Because of Lemma~2.1,
we obtain
\begin{equation}
[S^{(1)}-2^{n-2}(B_{0}^{(1)}+B_{1}^{(1)})]
-[S-2^{n-2}(a_{0}+a_{1})]
>\epsilon^{(0)}>0,
\end{equation}
where $S^{(1)}$ is a sum of all coefficients of $|\Psi^{(1)}\ket$.
Then, we assume
$S^{(1)}-2^{n-2}(B_{0}^{(1)}+B_{1}^{(1)})<0$.
After applying $(R_{\pi}D)$ on $|\Psi^{(1)}\ket$,
we get $|\Psi^{(2)}\ket$
specified by
$0<B_{0}^{(2)}<B_{1}^{(2)}<B_{j}^{(2)}$
for $2(l+m)\leq j\leq 2^{n}-1$.
Because of Lemma~2.2,
we get
\begin{equation}
[S^{(2)}-2^{n-2}(B_{0}^{(2)}+B_{1}^{(2)})]
-[S^{(1)}-2^{n-2}(B_{0}^{(1)}+B_{1}^{(1)})]
>\epsilon^{(1)}>\epsilon^{(0)}>0.
\end{equation}
Consequently,
if $S^{(k)}-2^{n-2}(B_{0}^{(k)}+B_{1}^{(k)})<0$,
$[S^{(k+1)}-2^{n-2}(B_{0}^{(k+1)}+B_{1}^{(k+1)})]$
increases by $\epsilon^{(0)}(>0)$
at least.
$k$ stands for the number of the $(R_{\pi}D)$ transformations
applied on the state
and $S^{(0)}=S$, $B_{0}^{(0)}=a_{0}$, $B_{1}^{(0)}=a_{1}$.
Because $\epsilon^{(0)}$ is defined by $\{a_{0},a_{1}\}$ and $l$,
$\epsilon^{(0)}$ is a definite finite value and positive.
Repeating the $(R_{\pi}D)$ finite times,
we can certainly make $[S^{(k)}-2^{n-2}(B_{0}^{(k)}+B_{1}^{(k)})]$
be nonnegative.

From Eq.~(\ref{a-coefficient}) and Eq.~(\ref{CoefficientsB1}),
during the $(R_{\pi}D)$ iteration,
we find that the phase shift is applied on the same basis vectors.
Therefore, the $(R_{\pi}D)$ iteration can be understood
as the inversion of Grover's iteration.
We use Grover's iteration
for enhancing an amplitude of a certain basis.

If $[S^{(k)}-2^{n-2}(B_{0}^{(k)}+B_{1}^{(k)})]$
comes to be nonnegative,
we start to do the $(\tilde{R}DR)$ operation again.
Using the $(\tilde{R}DR)$ operation and 
the $(R_{\pi}D)$ iteration,
we can always transform $|\psi_{n}\ket$ to the uniform superposition.
How many times do we need to apply $(R_{\pi}D)$ on a state
to obtain the relation,
$S^{(k)}-2^{n-2}(B_{0}^{(k)}+B_{1}^{(k)})\geq 0$?

Estimating it,
first,
we introduce notations,
\beqa
\epsilon^{(k)}
&=&
(2l-2^{n-1})B_{0}^{(k)}+(2^{n}-2l)B_{1}^{(k)}, \non \\
{\cal F}^{(k)}
&=&
S^{(k)}-2^{n-2}(B_{0}^{(k)}+B_{1}^{(k)}).
\eeqa
Because of Lemma~2,
if ${\cal F}^{(k)}<0$,
we get relations,
\beqa
{\cal F}^{(k+1)}-{\cal F}^{(k)}
&>&\epsilon^{(k)}>0, \non \\
\epsilon^{(k+1)}-\epsilon^{(k)}
&\geq&
-[(2^{n}-2l)/2^{n-2}]{\cal F}^{(k)}
\geq
-{\cal F}^{(k)}/2^{n-3}
>0,
\eeqa
where we give the minimum of $(2^{n}-2l)$ by $2$.
Consequently,
if
${\cal F}^{(0)}<{\cal F}^{(1)}<\cdots
<{\cal F}^{(K-1)}<{\cal F}^{(K)}<0$,
we estimate $\epsilon^{(k)}$
($k=1,2,\cdots$) recurrently,
\beqa
\epsilon^{(1)}
&\geq&
-x{\cal F}^{(0)}+\epsilon^{(0)}
>
-x{\cal F}^{(1)}
(>0), \non \\
\epsilon^{(2)}
&\geq&
-x{\cal F}^{(1)}+\epsilon^{(1)}
>
-2x{\cal F}^{(1)}
>
-2x{\cal F}^{(2)}
(>0), \non \\
\cdots \non \\
\epsilon^{(K)}
&\geq&
-x{\cal F}^{(K-1)}+\epsilon^{(K-1)}
>
-Kx{\cal F}^{(K-1)}
>
-Kx{\cal F}^{(K)}
(>0),
\eeqa
where $x=1/2^{n-3}$.
From these relations,
assuming $Kx\leq 1$,
we obtain
\beqa
(0>){\cal F}^{(2)}
&>&{\cal F}^{(1)}+\epsilon^{(1)}
>(1-x){\cal F}^{(1)}
>(1-x){\cal F}^{(0)}, \non \\
(0>){\cal F}^{(3)}
&>&{\cal F}^{(2)}+\epsilon^{(2)}
>(1-2x){\cal F}^{(2)}
>(1-x)(1-2x){\cal F}^{(0)}, \non \\
\cdots \non \\
{\cal F}^{(K+1)}
&>&{\cal F}^{(K)}+\epsilon^{(K)}
>(1-Kx){\cal F}^{(K)}
\geq\prod_{k=1}^{K}(1-kx){\cal F}^{(0)}.
\eeqa

If
${\cal F}^{(K+1)}+\epsilon^{(0)}\geq 0$,
we obtain ${\cal F}^{(K+2)}>0$
and
we can conclude we need to apply
the $(R_{\pi}D)$ transformation $(K+2)$ times at most.
To derive the upper bound on times
we have to apply
the $(R_{\pi}D)$ transformations,
we estimate
$\epsilon^{(0)}$
and ${\cal F}^{(0)}$,
\beqa
\epsilon^{(0)}
&\geq&
(2^{n}-2l)a_{1}\geq 2a_{1}, \non \\
{\cal F}^{(0)}
&>&
2^{n}a_{0}-2^{n-2}(a_{0}+a_{1})
\geq -2^{n-2}a_{1},
\eeqa
and we obtain
\begin{equation}
{\cal F}^{(K+1)}+\epsilon^{(0)}
\geq
\prod_{k=1}^{K}(1-kx){\cal F}^{(0)}
+\epsilon^{(0)}
\geq
-2^{n-2}a_{1}[\prod_{k=1}^{K}(1-kx)-x].
\end{equation}

Therefore,
to estimate the lower bound on
$K$ for
${\cal F}^{(K+1)}+\epsilon^{(0)}\geq 0$,
we have to derive
the lower bound on $K$
for the large $n$ (small $x$) limit, where
\begin{equation}
\prod_{k=1}^{K}(1-kx)\leq x
\quad\quad
\mbox{for $0<x \ll 1$}.
\end{equation}
Because
$\lim_{m\to +\infty}
[1-(1/m)]^{-m}=e(>2)$,
if $x_{0}$ is small enough,
we obtain
\begin{equation}
\prod_{k=\lceil \sqrt{1/x} \rceil}
^{2\lceil \sqrt{1/x} \rceil -1}
(1-kx)
<
(1-\frac{1}{\lceil \sqrt{1/x} \rceil})
^{\lceil \sqrt{1/x} \rceil}
<
\frac{1}{2},
\end{equation}
for $0<\forall x <x_{0}\ll 1$
($\lceil \sqrt{1/x} \rceil$ is the minimum integer
that does not below
$\sqrt{1/x}$).
Remembering $x=1/2^{n-3}$,
we get
\begin{equation}
x>
[
\prod_{k=\lceil \sqrt{1/x} \rceil}
^{2\lceil \sqrt{1/x} \rceil -1}
(1-kx)
]^{n-3}
>
\prod_{k=1}
^{(n-2)\lceil \sqrt{1/x} \rceil -1}
(1-kx).
\end{equation}
Consequently,
the lower bound on $K$ is
$(n-2)\lceil \sqrt{1/x} \rceil -1
\sim
O(n2^{n/2})$.
We have to apply the $(R_{\pi}D)$
transformation $O(n2^{n/2})$
times at most.
(See {\S}Appendix~B.)

Using Eq.~(\ref{CoefficientsB1}),
we can compute $\{B_{k}\}$ with $poly(n)$ steps
by classical computation,
because the number of different coefficients
in $\{B_{k}\}$ is
equal to $(\lfloor n/2 \rfloor +1)$.

%%%%%%%%%%%%%%%%%%%%%%%%%%%%%%%%%%%%%%%%%
\section{The whole procedure}
\lab{WholeProc}
In this section,
we show the whole procedure for building $|\psi_{n}\ket$
and give a sketch of implementation
for our procedure.
%%%%%%%%%%%%%%%%%%%%%%%%%%%%%%%%%%%%%%%%%%%%%%
We also show that we can use it
for building more general entangled states.
%%%%%%%%%%%%%%%%%%%%%%%%%%%%%%%%%%%%%%%%%%%%%%

As a result of discussion we have had,
we obtain the whole procedure to build
$|\psi_{n}\ket$ as follows.
(We describe the procedure reversely
from $|\psi_{n}\ket$ to the uniform superposition.
Throughout our procedure,
we take $\{|x\ket|x\in\{0,1\}^{n}\}$
as basis vectors.)
\begin{enumerate}
\item
%step1
We consider an $n$-qubit register
that is in the state of $|\psi_{n}\ket$
for an initial state.
(We assume all coefficients of basis vectors are
positive or equal to $0$.)
\item
%step2
If the state of the register is equal to the uniform superposition,
stop operations.
If it is not equal to the uniform superposition,
go to step~3.
\item
%step3
Let $a_{min}$ be the minimum coefficient for basis vectors
in the state of the register
and $a_{min+1}$ be the coefficient next to $a_{min}$.
Examine whether $a_{min}$ and $a_{min+1}$ satisfy
the sufficient condition
of Lemma~1
or not.
If they satisfy it,
carry out the $(\tilde{R}DR)$ operation,
shift the phases of 
basis vectors
which have negative coefficients by $\pi$,
and then go to step~2.
If they do not satisfy it,
go to step~4.
\item
%step4
Apply the $(R_{\pi}D)$ transformation on the register
and go to step~3.
\end{enumerate}

Before executing this procedure,
we need to trace a variation of coefficients of basis
in each step
by classical computation,
because
we have to know which basis vectors have
the coefficients $a_{min}$ and $a_{min+1}$,
find the phase shift parameter
of the $(\tilde{R}DR)$ operation, and so on.
From these results,
we construct a network of quantum gates.
The amount of classical computation is comparable
with the number of steps for the whole quantum transformations.

We now sketch out the points of networks
of quantum gates for our procedure.
Because it is a chain
of phase shift transformations
and Grover's operation $D$s,
we discuss the networks of quantum gates for them.

First, we discuss the phase shift transformation.
In the $(\tilde{R}DR)$ operation,
we shift the phases by $\theta$ on half of basis vectors
which have coefficients $a_{k}$ (as $a_{min+1}$)
and by $(-\theta)$ on the other half of them.
Constructing networks for $R_{\theta}$,
we prepare two registers and a unitary transformation
$U_{f}$,
\begin{equation}
|x\ket \otimes |y\ket
\stackrel{U_{f}}{\longrightarrow}
|x\ket \otimes |y\oplus f(x)\ket,
\lab{functionF}
\end{equation}
where the first (main) register is made from $n$ qubits,
the second (auxiliary) register is made from
$m=\lceil \log_{2}(n+1)\rceil$ qubits
initialized to $|0\cdots 0\ket$,
and
\begin{equation}
f(x)=
(\mbox{the number of ``$1$'' in the binary string of $x$}).
\lab{definition-fx-number1}
\end{equation}
Obtaining $f(x)$ on classical computation,
we need $O(nm)\sim O(n\log_{2} n)$ classical gates
(XOR, and so on)
and $O(m)\sim O(\log_{2} n)$
other auxiliary classical bits.
Therefore,
we can construct $U_{f}$ with $O(n\log_{2} n)$
elementary quantum gates
%%%%%%%%%%%%%%%%%%%%%%%%%%%%%%%%%%%%%%%%%%%%%%%%%%%%%%
(\cite{Feynman}\cite{Barenco}\cite{Bennett73}
and see {\S}Appendix~A).
%%%%%%%%%%%%%%%%%%%%%%%%%%%%%%%%%%%%%%%%%%%%%%%%%%%%%%

To execute the selective phase shift efficiently,
we apply it on the second register
instead of the first register.
Because the phase shift matrix defined in
Eq.~(\ref{matrix-phaseshift})
is diagonal,
we can do this way.
After shifting the phases,
we apply $U_{f}$ again and
initialize the second register.
Unnecessary entanglement between the first and the second register is removed.

To see these operations precisely,
we apply $U_{f}$ on $|\psi_{n}\ket$
defined in Eq.~(\ref{PsiNForm}).
We get
\beqa
U_{f}|\psi_{n}\ket\otimes|0\ket
&=&U_{f}
\sum_{k=0}^{\lfloor n/2 \rfloor} a_{k}|k\ket_{s}
\otimes|0\ket \non \\
&=&
\left\{
\begin{array}{ll}
\sum_{k=0}^{(n-1)/2} a_{k}
[|k)\otimes|k\ket+|n-k)\otimes|n-k\ket],
& \mbox{($n$ is odd)} \\
\sum_{k=0}^{(n/2)-1} a_{k}
[|k)\otimes|k\ket+|n-k)\otimes|n-k\ket] \non \\
\quad\quad\quad\quad\quad\quad
+a_{n/2}|n/2)\otimes|n/2\ket,
& \mbox{($n$ is even)}
\end{array}
\right.
\eeqa
where $|k)$ is an equally weighted superposition of
$k$ excited qubits
($|k\ket_{s}=|k)+|n-k)$
except for $|n/2\ket_{s}=|n/2)$
where $n$ is even).
We shift the phases of basis vectors
$|k\ket$, $|n-k\ket$ on the second register,
instead of $|k)$, $|n-k)$
(that contain
$2(^{n}_{k})$
binary states)
on the first register
(where $k\neq n/2$).
This implementation reduces the number of basis vectors
on which we apply the phase shift operation
from
$2(^{n}_{k})$
to $2$,
and we can save elementary quantum gates.
Using another auxiliary qubit,
we can carry out the phase shift with $O(\log_{2} n)$
elementary quantum gates
%%%%%%%%%%%%%%%%%%%%%%%%%%%%%%%%%%%%%%%%%%%%%%%%%%%
(\cite{Barenco}\cite{Cleve-Ekert} and see {\S}Appendix~A).
%%%%%%%%%%%%%%%%%%%%%%%%%%%%%%%%%%%%%%%%%%%%%%%%%%%
If $n$ is even and $k=n/2$,
we can't decide
which basis vectors we have to shift the phases
by $\theta$ or $(-\theta)$.
In this case, we refer to not only
$|n/2\ket$ on the second register
but also the first qubit of the first register
(cf. $|\psi_{4}\ket$ defined in Eq.~(\ref{Psi4Form})).

Then, we discuss how to construct the quantum
network of $D$.
It is known that $D$ can be decomposed to the form,
$D=-WRW$,
where $W=H^{(1)}\otimes \cdots \otimes H^{(n)}$
(the Walsh-Hadamard transformation on $n$ qubits
of the main register),
and $R$ is a phase shift by $\pi$
on $|0\cdots 0\ket$ of $n$ qubits\cite{Grover}.
$D$ takes $O(n)$ steps.

We repeat the $(R_{\pi}D)$ transformation
$O(n2^{n/2})$times at most
before the $(\tilde{R}DR)$ operation.
If we do the $(R_{\pi}D)$ iteration before every $(\tilde{R}DR)$,
we carry out it $\lfloor n/2 \rfloor$ times.
Therefore,
the $(R_{\pi}D)$ iterations take the main part of the whole steps.
Because $(R_{\pi}D)$ takes $O(n\log_{2} n)$ steps,
we need $O((n^{3}\log_{2} n)\times 2^{n/2})$
steps for the whole procedure in total
at most.

Finally,
we consider the case where all of $\{a_{k}\}$
are neither positive nor real.
Doing the selective phase shift on the basis vectors
with complex or  negative real
coefficients in $|\psi_{n}\ket$ to cancel the phases,
we obtain
a superposition whose all coefficients are real
and nonnegative.
After this operation,
we can apply our procedure on the state.

In our method,
we don't fully use the symmetry of $|\psi_{n}\ket$.
Essential points that we use are as follows.
First,
the number of basis vectors
that have the same coefficient is always even.
Second,
we can efficiently shift
the phase of half the basis vectors
that have the same coefficients.
Third,
the number of different coefficients $\{a_{k}\}$
is $poly(n)$.
Therefore,
we can apply our method to build more general entangled states
that have above properties.

%%%%%%%%%%%%%%%%%%%%%%%%%%%%%%%%%%%%%%%%%%%%%%%
\begin{figure}[ht]
\begin{center}
\includegraphics[scale=0.8]{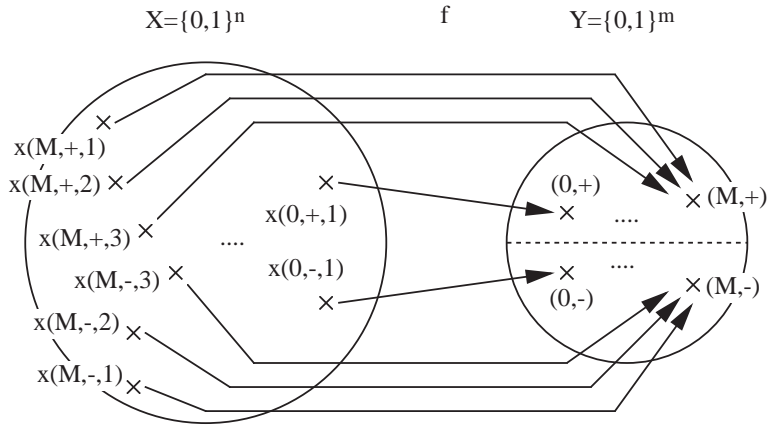}
\caption{A function $f$
defined in Eq.~(\ref{def-general-function-f}).}
\label{mapgenf2-epsf}
\end{center}
\end{figure}
Here, we discuss applying our method for building
more general entangled states than $|\psi_{n}\ket$.
We consider an entangled state
defined by a function $f$,
as follows,
\begin{equation}
f:X=\{0,1\}^{n}\rightarrow Y=\{0,1\}^{m},
\lab{GeneralFunction1}
\end{equation}
where $m=\lceil \log_{2}(M+1)\rceil +1$
and $M$ is polynomial in $n$.
We assume
we can label elements of image
caused by $f$ from $X=\{0,1\}^{n}$ by
$\{(0,\pm),(1,\pm),\cdots,(M,\pm)\}$.
We also assume
the number of $X$'s elements mapped to $(k,+)$
and
the number of them mapped to $(k,-)$
are equal to $l_{k}$ for $k=0,\cdots,M$,
where
$2\sum_{k=0}^{M}l_{k}=2^{n}$.
We can describe the function $f$ by
\begin{equation}
f(x(k,\epsilon,\zeta))=(k,\epsilon),
\lab{def-general-function-f}
\end{equation}
where $k=0,1,\cdots,M$,
and
$\epsilon=\pm$,
and $\zeta=1,\cdots,l_{k}$,
as shown in
Figure~\ref{mapgenf2-epsf}.

Then we consider the following
$n$-qubit partially entangled state,
\begin{equation}
|\Psi_{n}\ket=
\sum_{k=0}^{M}
\sum_{\epsilon=\pm}
\sum_{\zeta=1}^{l_{k}}
c_{k}|x(k,\epsilon,\zeta)\ket
\quad\quad
\mbox{for $n\geq 2$},
\lab{GeneralPartiallyEntangledStates}
\end{equation}
where $\{c_{k}\}$ are complex.
The number of sets of basis vectors classified by $\{c_{k}\}$
is $(M+1)$,
and the number of basis vectors
that have the coefficient $c_{k}$
is $2l_{k}$.

Executing the selective phase shift efficiently,
we apply $U_{f}$ of Eq.~(\ref{def-general-function-f})
to write $(k,\epsilon)$ on the $m$-qubit
second register
and apply phase shift transformation
on it.
We can shift the phase by $\theta$
or $(-\theta)$ according to $\epsilon$
in the $(\tilde{R}DR)$ operation.
To transform $|\Psi_{n}\ket$ to the uniform superposition,
we have to do the $(\tilde{R}DR)$ operation $M$ times.
Consequently, the $(R_{\pi}D)$ transformation is repeated
$M\times O(n2^{n/2})$
times at most.
It is desirable that $M$ is $poly(n)$.
%%%%%%%%%%%%%%%%%%%%%%%%%%%%%%%%%%%%%%%%%%%%%%%

%%%%%%%%%%%%%%%%%%%%%%%%%%%%%%%%%%%%%%%%%
\section{Discussion}
\lab{DISCUSSION}
It is known that any unitary transformation
$U$ $(\in \mbox{\boldmath $U$}(2^{n}))$
can be constructed from $O(n^{3}2^{2n})$ elementary gates
at most\cite{Barenco}.
In comparison with this most general case,
our method is efficient,
although the number of gates increases exponentially in $n$.

%%%%%%%%%%%%%%%%%%%%%%%%%%%%%%%%%%%%%%%%%%%%%%%%%%%%%%%%%%%%%
C.~H.~Bennett et al.
discuss transmitting
classical information via quantum noisy
%%%%%%%%%%%%%%%%%%%%%%%%%%%
\\
%%%%%%%%%%%%%%%%%%%%%%%%%%%
channels\cite{Bennett-Fuchs}.
It is shown when two transmissions of
the two-Pauli channel are used,
the optimal states for transmitting classical information are
partially entangled states of two qubits.
Therefore, we can expect our method is available for quantum communication.

Grover's algorithm was proposed
as a solution of the SAT(satisfiability) problems.
It finds a certain combination
from all of $2^{n}$ possible combinations of
$n$ binary variables.
From a different view,
what Grover's method does is
enhancing an amplitude of a certain basis vector
specified with an oracle
for a superposition of $2^{n}$ basis vectors.
%%%%%%%%%%%%%%%%%%%%%%%%%%%%%%%%%%%%%%%%%%%%%%%%%%%%%%%%%%%%%
In our method,
we use Grover's method for
adjusting amplitudes of basis vectors.

We can't show whether our procedure is optimal or not
in view of the number of elementary gates.
Because we don't use
the symmetry of $|\psi_{n}\ket$ enough,
our method seems to waste steps.

Recently,
constructing approximately an optimal state
for Ramsey
spectroscopy
by spin squeezing
has been proposed\cite{Ulam-Orgikh}.
This state also has symmetry like
Eq.~(\ref{PsiNForm}),
and 
it is characterized by one parameter.

%%%%%%%%%%%%%%%%%%%%%%%%%%%%%%%%%%%%%%%%%
\bigskip
\noindent
{\bf \large Acknowledgements}

We would like to thank Dr.~M.~Okuda
and
Prof.~A.~Hosoya for critical reading
and valuable comments.

%%%%%%%%%%%%%%%%%%%%%%%%%%%%%%%%%%%%%%%%%

%%%%%%%%%%%%%%%%%%%%%%%%%%%%%%%%%%%%%%%%%
\appendix
\section{Networks of quantum gates}
We construct networks of quantum gates
for our method concretely.
For notations of networks
and quantum gates,
we refer to A.~Barenco et al.\cite{Barenco}.

\subsection{The network of $U_{f}$}
$U_{f}$
(defined in
Eq.~(\ref{functionF}),
(\ref{definition-fx-number1}) or
Eq.~(\ref{def-general-function-f}))
is given by the controlled gate,
which causes the unitary transformation
on the second register
under the value of the first register.
Constructing the controlled gate of $U_{f}$
with $poly(n)$ quantum elementary gates,
we can use our method efficiently.

We consider a network
for $U_{f}$ defined in
Eq.~(\ref{functionF}) and
Eq.~(\ref{definition-fx-number1}).
$f(x)$ represents the number of ``$1$'' bit
in the binary string $x$.
Writing the first (main) and second (auxiliary) register by
$|X_{n},X_{n-1},\cdots,X_{2},X_{1}\ket\otimes|S\ket$,
where $|S\ket$ is
made up of $m=\lceil \log_{2}(n+1)\rceil$
qubits and
initialized to $|0\cdots 0\ket$,
we can write the quantum networks
as the following program.
(For the notation of the program,
we referred to Cleve et al.\cite{Cleve-DiVincenzo}.)
\begin{tabbing}
{\bf Program} adder-1 \\
\quad \={\bf quantum registers}: \\
\>$X_{1},X_{2},\cdots,X_{n}$:  qubit registers \\
\>$S$: an m-qubit register \\
{\bf for} $k=1$ {\bf to} $n$ {\bf do} \\
\quad\quad\=
$S\leftarrow (S+X_{k}) \quad\mbox{mod}\;2^{m}$. \\
\end{tabbing}

\begin{figure}[ht]
\begin{center}
\includegraphics[scale=0.8]{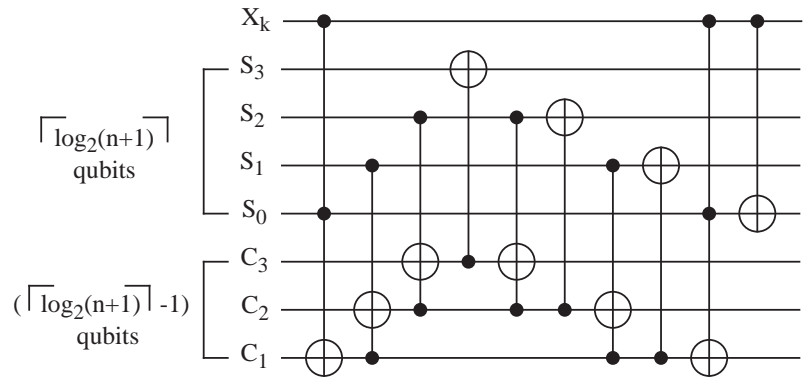}
\caption{The network of the adder-2
for $m=\lceil \log_{2}(n+1)\rceil =4$.}
\label{adder2d-epsf}
\end{center}
\end{figure}
To write a program for the addition
of $X_{k}$ in the adder-1,
we describe the qubits of the second register
by $|S_{m-1},\cdots,S_{1},S_{0}\ket$,
introduce other auxiliary
qubits $|C_{m-1},\cdots,C_{1}\ket$,
and
use $C_{j}$ as a carry bit of
addition at the $(j-1)$th bit.
We can write the program
as follows.
\begin{tabbing}
{\bf Program} adder-2 \\
\quad \={\bf quantum registers}: \\
\>$X_{k}$: a qubit register \\
\>$S_{0},S_{1},\cdots,S_{m-1}$: qubit registers \\
\>$C_{1},C_{2},\cdots,C_{m-1}$: auxiliary qubit registers
(initialized and finalized to 0) \\
$C_{1}\leftarrow C_{1}\oplus(S_{0}\land X_{k})$ \\
{\bf for} $j=2$ {\bf to} $m-1$ {\bf do} \\
\>\quad \=$C_{j}\leftarrow C_{j}\oplus(C_{j-1}\land S_{j-1})$ \\
{\bf for} $j=m-1$ {\bf down} {\bf to} $2$ {\bf do} \\
\>\> $S_{j}\leftarrow S_{j}\oplus C_{j}$ \\
\>\> $C_{j}\leftarrow C_{j}\oplus(C_{j-1}\land S_{j-1})$ \\
$S_{1}\leftarrow S_{1}\oplus C_{1}$ \\
$C_{1}\leftarrow C_{1}\oplus(S_{0}\land X_{k})$ \\
$S_{0}\leftarrow S_{0}\oplus X_{k}$. \\
\end{tabbing}

Because we don't use interference,
we can describe these operations with a higher
level language of classical computation.
In this program,
to avoid obtaining unnecessary entanglement,
we initialize and finalize all
auxiliary qubits $\{|C_{j}\ket\}$
to $|0\ket$.
Figure~\ref{adder2d-epsf} shows a network of
this program for $m=4$.
Repeating the quantum network
of the adder-2
for each $X_{k}$ $(k=1,\cdots,n)$,
we can construct the adder-1.

We estimate
the number of
quantum elementary gates
to construct the
adder-1.
In Figure~\ref{adder2d-epsf}, we use
$2(\lceil \log_{2}(n+1) \rceil -1)$ Toffoli gates
(that maps
$|x,y,z\ket\rightarrow|x,y,z\oplus(x\wedge y)\ket$)
and
$\lceil \log_{2}(n+1) \rceil$ controlled-NOT gates
for the adder-2.
Because we repeat the adder-2 $n$ times,
the number of the whole steps for the adder-1
is equal to $n(3\lceil \log_{2}(n+1) \rceil -2)$.

\subsection{Construction of $\bigwedge_{n}(R_{z}(\alpha))$}
From now on,
we often use a
$\bigwedge_{n}(R_{z}(\alpha))$
gate,
where
$R_{z}(\alpha)$
is given in the form,
\begin{equation}
R_{z}(\alpha)
=\exp(i\alpha\sigma_{z}/2)
=
\left[
\begin{array}{cc}
\exp(i\alpha/2) & 0 \\
0 & \exp(-i\alpha/2)
\end{array}
\right].
\end{equation}
(We describe the $\mbox{controlled}^{m}\mbox{-}U$
by $\bigwedge_{m}(U)$,
where $\forall U\in \mbox{\boldmath $U$}(2)$.
$\bigwedge_{m}(U)$
has an $m$-qubit control subsystem and
a one-qubit target subsystem.
It works as follows.
If all $m$ qubits of control subsystem are equal to $|1\ket$,
$\bigwedge_{m}(U)$ applies $U$ on a target qubit.
Otherwise
$\bigwedge_{m}(U)$ does nothing.
We can write the Toffoli gate by $\bigwedge_{2}(\sigma_{x})$,
the controlled-NOT gate by $\bigwedge_{1}(\sigma_{x})$,
and any $\mbox{\boldmath $U$}(2)$ gate for one qubit
by $\bigwedge_{0}$.)
Here,
we consider how to construct
it
from elementary gates.

\begin{figure}[ht]
\begin{center}
\includegraphics[scale=0.8]{lnrza.epsf}
\caption{Decomposition of a $\bigwedge_{n}(R_{z}(\alpha))$ gate.}
\label{lnrza-epsf}
\end{center}
\end{figure}
\begin{figure}[ht]
\begin{center}
\includegraphics[scale=0.8]{l1rzb.epsf}
\caption{Decomposition of a $\bigwedge_{1}(R_{z}(\beta))$ gate.}
\label{l1rzb-epsf}
\end{center}
\end{figure}
At first, using relations,
\[
R_{z}(\alpha/2)\sigma_{x}R_{z}(-\alpha/2)\sigma_{x}
=R_{z}(\alpha),
\quad\quad
\mbox{and}
\quad\quad
R_{z}(\alpha/2)R_{z}(-\alpha/2)
=\mbox{\boldmath $I$},
\]
we can decompose a $\bigwedge_{n}(R_{z}(\alpha))$ gate
to a $\bigwedge_{1}(R_{z}(\alpha/2))$ gate,
a $\bigwedge_{1}(R_{z}(-\alpha/2))$ gate
and two $\bigwedge_{n-1}(\sigma_{x})$ gates,
as shown in Figure~\ref{lnrza-epsf}.
Seeing Figure~\ref{l1rzb-epsf},
we can decompose a $\bigwedge_{1}(R_{z}(\beta))$ gate
to an $R_{z}(\beta/2)$ gate, an $R_{z}(-\beta/2)$ gate
and two controlled-NOT gates.
We have to only consider
how to make a $\bigwedge_{n-1}(\sigma_{x})$ gate
from elementary gates on an $(n+1)$-qubit network.
Especially, we pay attention to the fact that
there is a qubit which is not used
by the $\bigwedge_{n-1}(\sigma_{x})$ gate
on the network.

It is shown that,
on an $(n+1)$-qubit network,
where $n\geq 6$,
a $\bigwedge_{n-1}(\sigma_{x})$ gate
can be decomposed to $8(n-4)$ Toffoli gates\cite{Barenco}.
Consequently, on the $(n+1)$-qubit network$(n\geq 6)$,
a $\bigwedge_{n}(R_{z}(\alpha))$ gate can be decomposed
to $16(n-4)$ Toffoli gates,
four $\bigwedge_{1}(\sigma_{x})$ gates and 
four $\bigwedge_{0}$ gates.
Therefore, $\bigwedge_{n}(R_{z}(\alpha))$
takes $8(2n-7)$ quantum elementary gates in total.

\subsection{The phase shift on certain basis vectors}
\begin{figure}[ht]
\begin{center}
\includegraphics[scale=0.8]{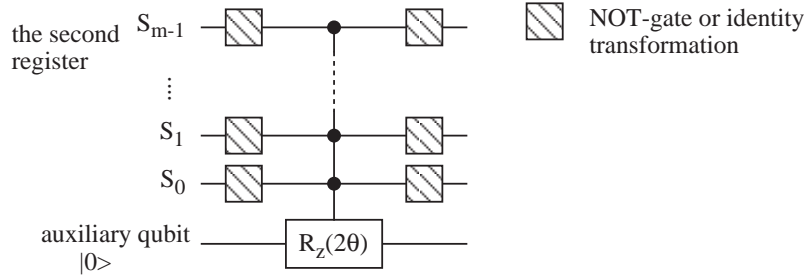}
\caption{The network of the phase shift on the second register.}
\label{rot2nd2-epsf}
\end{center}
\end{figure}

Figure~\ref{rot2nd2-epsf} shows a quantum network
for the selective phase shift by $\theta$
on a certain basis vector of the second register
defined in Eq.~(\ref{functionF}).
In Figure~\ref{rot2nd2-epsf}, we use a
$\bigwedge_{m}(R_{z}(2\theta))$
gate.
Setting the auxiliary qubit being $|0\ket$,
$\bigwedge_{m}(R_{z}(2\theta))$ generates an eigenvalue
$\exp(i\theta)$
if and only if the second register is in the state
$|1\cdots 1\ket$.
This technique is called ``kick back''\cite{Cleve-Ekert}.

In Figure~\ref{rot2nd2-epsf},
a shaded box stands for the NOT-gate given by $\sigma_{x}$
($|0\ket\rightarrow|1\ket$,
$|1\ket\rightarrow|0\ket$)
or the identity transformation.
Deciding which gates are set in each shaded box,
$\sigma_{x}$ or {\boldmath $I$},
we can select a basis vector on which we shift the phase.

In case $m\geq 6$,
it has been already shown that
a $\bigwedge_{m}(R_{z}(2\theta))$ gate can be constructed
from $8(2m-7)$ quantum elementary gates at most.
Seeing Figure~\ref{rot2nd2-epsf},
we find that the selective phase shift
on the second register takes
$2m+8(2m-7)=2(9m-28)\sim O(m)$ gates at most.
Building $|\psi_{n}\ket$,
we can carry out the phase shift
on certain basis vectors
on the second register with $O(\log_{2}n)$ steps.

\subsection{The network of $D$}
\begin{figure}[ht]
\begin{center}
\includegraphics[scale=0.8]{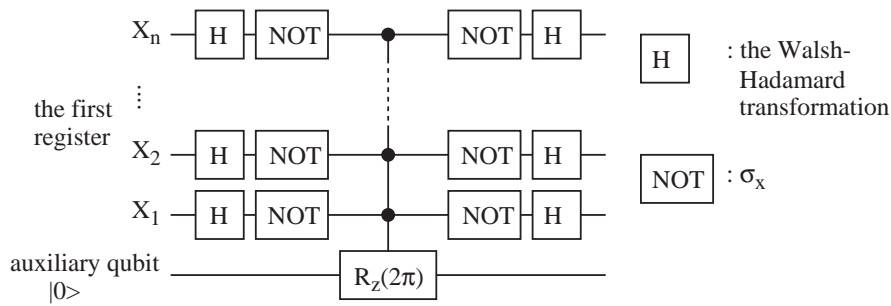}
\caption{The network of $D$.}
\label{netDd-epsf}
\end{center}
\end{figure}

Figure~\ref{netDd-epsf} shows a network of $D$.
Since this network consists of
$4n$ $\bigwedge_{0}$ gates
and a $\bigwedge_{n}(R_{z}(2\pi))$ gate,
it takes $4(5n-14)$ elementary gates, in case $n \geq 6$.
Therefore, $D$ takes $O(n)$ steps.

\subsection{Estimation of steps}
How many elementary gates
do we need to construct
$|\psi_{n}\ket$
defined in Eq.~(\ref{PsiNForm}) or
$|\Psi_{n}\ket$
defined in Eq.~(\ref{GeneralPartiallyEntangledStates})
from the uniform superposition?
If $M$ (the number of sets of basis vectors
classified by their coefficients) is $poly(n)$,
and if the function $U_{f}$
defined in Eq.~(\ref{functionF}) can be constructed
from $poly(n)$ elementary gates,
the $(R_{\pi}D)$ iterations take
the main part of the whole steps.

In the $(R_{\pi}D)$ transformation,
we do the following operations.
Applying $D$ on the $n$-qubit first register,
preparing the initialized $m$-qubit second register,
we apply $U_{f}$ on both of the registers
as Eq.~(\ref{functionF}).
Then,
we shift the phases of basis vectors
on the second register.
Finally,
we apply $U_{f}$ again to initialize the second register.
It has been already shown $D$ takes $4(5n-14)$ steps,
where $n \geq 6$.
The number of steps that a network of $U_{f}$ takes
depends on the function $f$.
For instance,
when we build $|\psi_{n}\ket$,
$U_{f}$ needs $O(n\log_{2} n)$ steps.

\begin{figure}[ht]
\begin{center}
\includegraphics[scale=0.8]{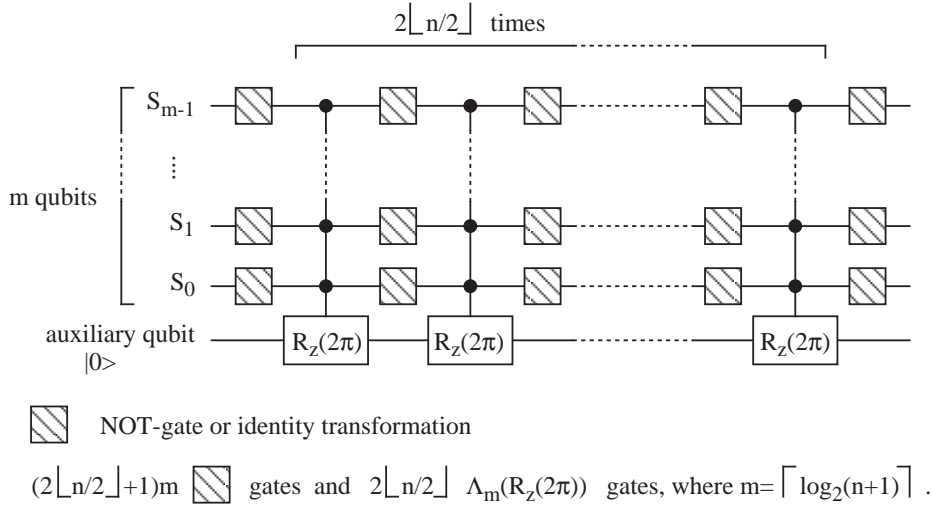}
\caption{The network of the phase shift
by $\pi$
on the second register for $R_{\pi}$}
\label{netRpi2-epsf}
\end{center}
\end{figure}
Figure~\ref{netRpi2-epsf} shows a network
of the phase shift by $\pi$
on the second register
with negative coefficients for $R_{\pi}$,
in the case of building $|\psi_{n}\ket$.
To inverse signs of negative coefficients,
we shift the phase at most
for $\lfloor n/2 \rfloor$
sets of basis vectors
characterized by coefficients.
Therefore,
we shift the phases of
$2\lfloor n/2 \rfloor$
basis vectors of the second register at most.
As a result,
the network can be constructed from
$(2\lfloor n/2 \rfloor +1)
\cdot
\lceil \log_{2}(n+1) \rceil$
$\bigwedge_{0}$ gates and
$2\lfloor n/2 \rfloor$
$\bigwedge_{\lceil \log_{2}(n+1) \rceil}(R_{z}(2\pi))$ gates.
We can carry out $R_{\pi}$ with $O(n\log_{2}n)$ steps.
Similarly,
in the case of $|\Psi_{n}\ket$,
we can estimate $R_{\pi}$
takes $O(M\log_{2}M)$ steps.

Building $|\psi_{n}\ket$,
we repeat the $(R_{\pi}D)$ transformation
$O(n2^{n/2})$times at most
before the $(\tilde{R}DR)$ operation.
If we do the $(R_{\pi}D)$ iteration before every $(\tilde{R}DR)$,
we carry out it $\lfloor n/2 \rfloor$ times.
Consequently,
we need $O((n^{3}\log_{2} n)\times 2^{n/2})$
steps for the whole procedure in total
at most.

%%%%%%%%%%%%%%%%%%%%%%%%%%%%%%%%%%%%%%%%%
\section{A variation of coefficients
during the $(R_{\pi}D)$ iteration}
We explicitly derive
a variation of coefficients
during the $(R_{\pi}D)$ iteration
for the case
described in (\ref{SimpleModelForRpiDIteration}),
and estimate how many times do we need
to apply $(R_{\pi}D)$
to make $[S^{(k)}-2^{n-2}(B_{0}^{(k)}+B_{1}^{(k)})]$
be nonnegative.
We find it takes $O(2^{n/2})$ times,
in spite of the results,
$O(n2^{n/2})$ times,
in \S\ref{CaseWhSufCondNotSat}.

Applying $(R_{\pi}D)$ on $|\Psi\ket$ defined by
(\ref{SimpleModelForRpiDIteration}),
we obtain
$R_{\pi}D|\Psi\ket=[B_{0}^{(1)},\cdots,B_{1}^{(1)},\cdots]$,
where
\beq
\left \{
\begin{array}{lll}
2^{n-1}B_{0}^{(1)}&=S-2^{n-1}a_{0}
&=(2^{n-1}-t)a_{0}+ta_{1}, \\
2^{n-1}B_{1}^{(1)}&=-S+2^{n-1}a_{1}
&=-(2^{n}-t)a_{0}+(2^{n-1}-t)a_{1}.
\end{array}
\right. 
\lab{SimpleModelRpiDCoefficient}
\eeq
Referring to \cite{Boyer-Brassard},
we write $t$ as
\beq
\sin^{2}\theta=\frac{t}{2^{n}},
\quad
(\cos^{2}\theta=\frac{2^{n}-t}{2^{n}}),
\lab{SimpleModelRpiDParat}
\eeq
where $0<\theta<(\pi/2)$,
and write $\{a_{0},a_{1}\}$ as
\beq
a_{0}=\frac{\sin\alpha}{\sqrt{2^{n}-t}},
\quad
a_{1}=\frac{\cos\alpha}{\sqrt{t}},
\lab{SimpleModelForRpiDIteration2}
\eeq
where $0\leq\alpha<(\pi/2)$.
Using 
(\ref{SimpleModelRpiDCoefficient}),
(\ref{SimpleModelRpiDParat}),
and (\ref{SimpleModelForRpiDIteration2}),
we can describe $\{B_{0}^{(1)},B_{1}^{(1)}\}$ by
\beq
\left \{
\begin{array}{lll}
B_{0}^{(1)}
&=(1/\sqrt{2^{n}})
[\cos 2\theta(\sin\alpha/\cos\theta)
+2\sin\theta\cos\alpha]
&=\sin(\alpha+2\theta)/\sqrt{2^{n}-t}, \\
B_{1}^{(1)}
&=(1/\sqrt{2^{n}})
[-2\cos\theta\sin\alpha
+\cos 2\theta(\cos\alpha/\sin\theta)]
&=\cos(\alpha+2\theta)/\sqrt{t}.
\end{array}
\right . 
\eeq

Writing coefficients of the state
on which $(R_{\pi}D)$ has been applied $k$ times
as $B_{0}^{(k)}$ and $B_{1}^{(k)}$,
we obtain
\beq
B_{0}^{(k)}=\frac{\sin(\alpha+2k\theta)}{\sqrt{2^{n}-t}},
\quad\quad
B_{1}^{(k)}=\frac{\cos(\alpha+2k\theta)}{\sqrt{t}}
\quad\quad
\mbox{for $k=0,1,2,\cdots,$}
\lab{explicit-Bs}
\eeq
where $B_{0}^{(0)}=a_{0}$, $B_{1}^{(0)}=a_{1}$.

Defining $S^{(k)}=(2^{n}-t)B_{0}^{(k)}+tB_{1}^{(k)}$,
we can derive
\beqa
&&S^{(k)}-2^{n-2}(B_{0}^{(k)}+B_{1}^{(k)}) \non \\
&=&(3\cdot 2^{n-2}-t)B_{0}^{(k)}+(t-2^{n-2})B_{1}^{(k)} \non \\
&=&\sqrt{2^{n}}
\{\sin[\alpha+(2k+1)\theta]
-\frac{1}{2\sin 2\theta}
\cos[\alpha+(2k-1)\theta]\} \non \\
&=&-\frac{\sqrt{2^{n-2}}}{\sin 2\theta}F^{(k)},
\lab{explicit-Sdash}
\eeqa
where
\beq
F^{(k)}=
\cos[\alpha+(2k+3)\theta].
\eeq
Since $0<\theta<(\pi/2)$ and $\sin 2\theta>0$,
it depends on a sign of $F^{(k)}$
whether $[S^{(k)}-2^{n-2}(B_{0}^{(k)}+B_{1}^{(k)})]$
is negative or not.
(With some calculations,
we can confirm that
(\ref{explicit-Bs}) and
(\ref{explicit-Sdash})
satisfy Lemma~2.)

Because of $0\leq \alpha <(\pi/2)$,
if $(2k+3)\theta=(\pi/2)$,
it is always accomplished that $F^{(k)}\leq 0$ and
$S^{(k)}-2^{n-2}(B_{0}^{(k)}+B_{1}^{(k)})\geq 0$.
Therefore, the number of times we need to apply
$(R_{\pi}D)$ doesn't
exceed $k_{MAX}$,
which is given as
\beq
k_{MAX}=\frac{1}{2\theta}(\frac{\pi}{2}-3\theta).
\eeq
On the other hand,
we can write $\theta$ as $\sin\theta=\sqrt{t/2^{n}}$
from (\ref{SimpleModelRpiDParat}),
and the minimum value of $t$ is $2$.
Taking the limit that
$t\sim O(1)$ and $n$ is large enough,
we obtain a relation,
$\sin\theta\sim\theta\sim\sqrt{t/2^{n}}$
and
\beq
k_{MAX}\sim\frac{\pi}{4}\sqrt{\frac{2^{n}}{t}}
\sim O(2^{n/2}).
\eeq
The $(R_{\pi}D)$ transformation is repeated
$O(2^{n/2})$ times at most to make
$[S^{(k)}-2^{n-2}(B_{0}^{(k)}+B_{1}^{(k)})]$
be nonnegative.

%%%%%%%%%%%%%%%%%%%%%%%%%%%%%%%%%%%%%%%%%

\end{document}